\documentclass{appolb}
\usepackage{epsfig}
\usepackage{color}
\usepackage{soul}
\usepackage{amsmath}
\usepackage{amssymb}
\newcommand {\bc}{\begin{center}}
\newcommand {\ec}{\end{center}}
\newcommand {\bea}{\begin{eqnarray}}
\newcommand {\eea}{\end{eqnarray}}
\newcommand {\be}{\begin{equation}}
\newcommand {\ee}{\end{equation}}

\def\lsim{\mathrel{\rlap{\lower4pt\hbox{\hskip1pt$\sim$}}
    \raise1pt\hbox{$<$}}}
\def\gsim{\mathrel{\rlap{\lower4pt\hbox{\hskip1pt$\sim$}}
    \raise1pt\hbox{$>$}}}

\begin{document}

\title{From cold Fermi fluids to the hot QGP%
\thanks{Presented by M.~Bluhm at "Critical Point and Onset of Deconfinement (CPOD) 2016", Wroc\l{}aw, Poland,
May 30th - June 4th, 2016}
}
\author{Marcus Bluhm$^{\,1,\,2}$ and Thomas Sch\"afer$^{\,2}$
\address{$^1$Institute of Theoretical Physics, University of Wroc\l{}aw, 50204 Wroc\l{}aw, Poland\\
$^2$Physics Department, North Carolina State University, Raleigh, NC 27695, USA}
}

\maketitle

\begin{abstract}
Strongly coupled quantum fluids are found in different forms, including ultracold Fermi gases or tiny droplets 
of extremely hot Quark-Gluon Plasma. Although the systems differ in temperature by many orders of magnitude, 
they exhibit a similar almost inviscid fluid dynamical behavior. In this work, we summarize some of the recent 
theoretical developments toward better understanding this property in cold Fermi gases at and near unitarity. 
\end{abstract}


\section{Introduction}%

Strongly coupled quantum fluids are studied with high interest in recent years~\cite{Bloch:2008zzb,Adams:2012th}. 
Despite apparant differences in temperature $T$, density $n$, or underlying microscopic theory, various quantum 
fluids share interesting features. For example, similar nearly inviscid flow was observed in both cold Fermi 
gases~\cite{O'Hara:2002zz} and the Quark-Gluon Plasma (QGP)~\cite{Romatschke:2007mq}. Based on data analyses, the 
shear viscosity $\eta$ to entropy density $s$ ratios were concluded to be as small as a few times the conjectured 
holographic KSS-limit~\cite{Kovtun03}. 

The QGP as color-deconfined state of hot and dense QCD matter is transciently formed in the expansion stage of 
high-energy heavy-ion collisions. Its properties can only indirectly be inferred from experimental data as only 
color-confined hadrons are measured in the detectors. In early attempts of describing the expanding matter with 
relativistic fluid dynamics~\cite{Romatschke:2007mq} evolution-averaged values for the QGP shear viscosity were 
extracted. Recent efforts~\cite{Bernhard:2016tnd,Karpenko:2015xea} aim instead at determining $\eta$ locally in 
dependence of $T$ and $n$. 

In cold Fermi gases, one can control the dominant $s$-wave interaction among atoms by altering an external magnetic 
field. When the interaction range becomes large compared to the interparticle spacing, the system is tuned into a 
Feshbach resonant state and the scattering cross section is bound only by unitarity. In this limit, the system is 
scale and conformally invariant and the matter properties are universal functions of $\,T$ and $n$~\cite{Ku2012}. At 
low $T$, this matter forms a superfluid. On different sides of the resonance, a Bose-Einstein condensate (BEC) of 
strongly bound molecules or a BCS-type atomic superfluid is realized with a smooth crossover in 
between~\cite{Bloch:2008zzb}. With increasing $T$, the matter undergoes a phase transition to a dense normal fluid 
whose collective behavior is correctly described by non-relativistic fluid dynamics~\cite{Schaefer:2014awa}. In the 
dilute gas limit, kinetic theory becomes applicable even in the unitarity limit. 

One way to extract $\eta$ in cold Fermi gases is to analyze expansion measurements of gas 
clouds~\cite{Cao:2010wa,Elliott:2014nn,Joseph:2014}. In the experiments, the gas is cooled and trapped in a harmonic 
optical potential, which determines the initial properties of the cloud, and atomic interactions are controlled 
magnetically. After release from the trapping potential the gas expands and its time-evolution can directly be 
observed. Most of the previous analyses concentrated on deducing trap-averaged values for $\eta$, however, and only 
recently a local determination became possible. 

\section{Anisotropic fluid dynamics for cold Fermi gases}%

The natural approach for studying the expansion dynamics of the dense gas cloud is Navier-Stokes fluid dynamics. 
In the dilute corona, where scatterings are rare and the system expands ballistically, the theory becomes however 
inapplicable and leads to unphysical results. To overcome this situation one needs an approach which is capable 
of describing locally the transition from fluid dynamical to free streaming behavior. Considering the full 
Boltzmann equation~\cite{Pantel:2014jfa} is one possibility, as free streaming is contained in the theory and the 
fluid dynamical limit is reproduced in the dense regime. Determining locally the temperature and density dependence 
of the shear viscosity within this approach is, nonetheless, a difficult task. 

Anisotropic fluid dynamics represents an alternative method. In contrast to Navier-Stokes fluid dynamics this theory 
can be derived from moments of the Boltzmann equation connected to an anisotropic distribution function. The form of 
this function is motivated by exact solutions of the kinetic theory in the free streaming limit for a harmonic trap 
potential~\cite{Menotti:2002,Pedri:2003,Chiacchiera:2013}. Assuming that the symmetry axes of the potential are 
aligned with the cartesian coordinates $a$ simplifies the set of evolution equations, see~\cite{Bluhm:2015raa} for 
details. In Lagrangian form, the continuity equation reads 
\be
\label{rho_lag}
 \left(\partial_t + \vec{u}\cdot\vec{\nabla}\right) \rho = -\rho \vec{\nabla}\cdot \vec{u}
\ee
and the conservation equations for momentum and energy read 
\bea
\label{u_lag}
 \left(\partial_t + \vec{u}\cdot\vec{\nabla}\right) u_i & = & - \frac{1}{\rho} \left( \nabla_i P 
 + \nabla^j \delta \Pi_{ij} \right) \, , \\
\label{e_lag}
 \left(\partial_t + \vec{u}\cdot\vec{\nabla}\right) \epsilon & = & - \frac{1}{\rho} \nabla^i \left( u_i P 
 + u^j \delta \Pi_{ij} \right) \, . 
\eea
Here, $\rho=mn$ is the mass density, $\vec{u}$ the fluid velocity, $P$ the pressure and $\epsilon={\cal E}/\rho$ the 
energy per mass. For a scale invariant Fermi gas at unitarity, pressure and energy density in the fluid rest frame 
are related via $P=\frac23{\cal E}_0$ with ${\cal E}_0={\cal E}-\frac12\rho\vec{u}^2$. The dissipative stress tensor 
$\delta \Pi_{ij}$ has a simple diagonal form, $\delta \Pi_{ij}=\sum_a\delta_{ia}\delta_{ja}\Delta P_a$, where 
$\Delta P_a=P_a-P$ for the anisotropic pressure components $P_a$. The latter are related to the anisotropic energy 
density components ${\cal E}_a$ via $P_a=2{\cal E}_a^0$ where ${\cal E}_a^0={\cal E}_a-\frac12\rho u_a^2$ and 
${\cal E}=\sum_a{\cal E}_a$. 

In anisotropic fluid dynamics, one treats these components as additional fluid dynamical variables obeying for given 
$a$ the evolution equation 
\be 
\label{e_a_lag}
 \left(\partial_t + \vec{u}\cdot\vec{\nabla}\right) \epsilon_a = - \frac{1}{\rho}
 \nabla_a \left[ u_a P + u_a \Delta P_a \right] 
 - \frac{P}{2\eta\rho}\Delta P_a \, .
\ee
By summing over $a$, the equation of energy conservation Eq.~(\ref{e_lag}) is recovered, i.e.~only two of the 
three components ${\cal E}_a=\rho\epsilon_a$ are independent. Moreover, Eq.~(\ref{e_a_lag}) represents a relaxation 
equation for the dissipative stresses $\Delta P_a$: for small $\eta/P$, one finds 
$\Delta P_a = - \eta\sigma_{aa} + {\cal O}\left((\eta/P)^2\right)$ and anisotropic fluid dynamics reduces to the 
Navier-Stokes theory. For large $\eta/P$, instead, Eq.~(\ref{e_a_lag}) becomes a conservation equation for the 
individual components ${\cal E}_a$. Anisotropic fluid dynamics provides, thus, a theoretical framework which, 
irrespective of the functional form of $\eta$, smoothly combines viscous fluid dynamical behavior in dense regions 
with ballistic expansion in the corona. 

Relativistic anisotropic fluid dynamics was developed to achieve a more reliable description of the dynamics in a 
heavy-ion collision~\cite{Martinez:2010sc,Florkowski:2010cf,Strickland:2014pga}. At early times, during the 
pre-equilibrium evolution, large momentum-space anisotropies build up. These manifest in sizeable differences between 
the longitudinal and transverse pressure components and in large viscous corrections which become even larger with 
increasing $\eta/s$ and in the low-$T$ regions of the fireball. Anisotropic fluid dynamics is able to handle such 
corrections, thus, extending the regime of applicability of fluid dynamics to situations far from isotropic thermal 
equilibrium. 

\section{Determination of $\eta$ from high-$T$ expansion data}%

The evolution equations of anisotropic fluid dynamics can be solved with standard numerical techniques. Based on the 
PPMLR scheme of Colella and Woodward~\cite{Colella:1984} implemented in~\cite{Blondin:1993,Schafer:2010dv}, we 
developed a code~\cite{Bluhm:2015raa} that solves Eqs.~(\ref{rho_lag})~-~(\ref{e_a_lag}) in the unitarity limit for 
given initial state and functional dependence of $\eta$. The initial state is isothermal. The density profile is 
sensitive to the exact dependence of $P$ on $T$ and $n$ and determined by the hydrostatic equilibrium equation 
$\vec{\nabla}P=-n\vec{\nabla}V$ for trap potential $V$. In the high-$T$ limit, we have $P=nT$ and the initial density 
has a simple Gaussian shape. After release from the trap, the system is governed by the evolution equations which are 
sensitive to $P({\cal E}^0)$ only. To ensure stability of the relaxation equation~(\ref{e_a_lag}), the simulation 
time-step has to be smaller than the minimal $\eta/P$ in the gas cloud. In~\cite{Bluhm:2015bzi}, we showed that the 
code is capable of perfectly reproducing the numerical results~\cite{Pantel:2014jfa} for solving the Boltzmann 
equation in the dilute gas limit. 

\begin{figure}[t]
\bc
 \includegraphics[width=0.33\textwidth]{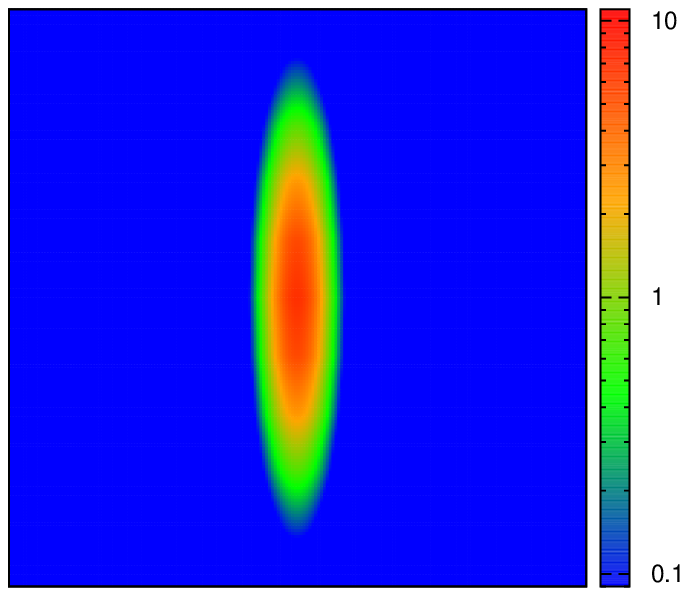}
 \hspace{-8mm}
 \includegraphics[width=0.33\textwidth]{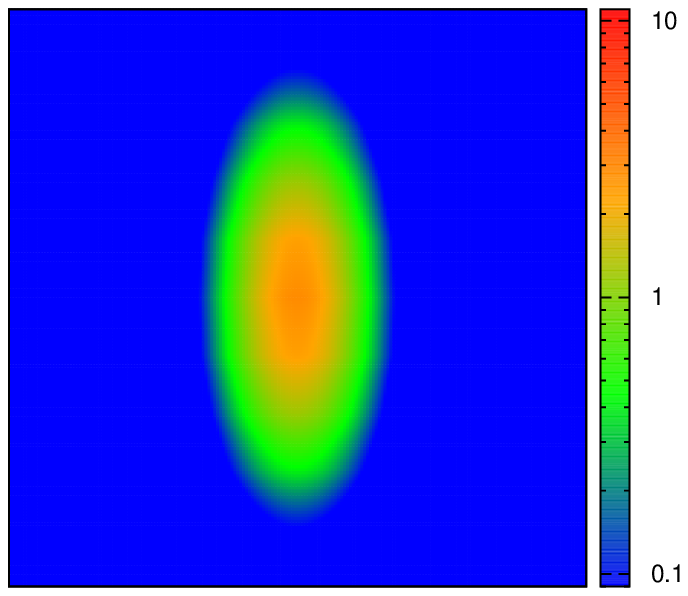}
 \hspace{-8mm}
 \includegraphics[width=0.33\textwidth]{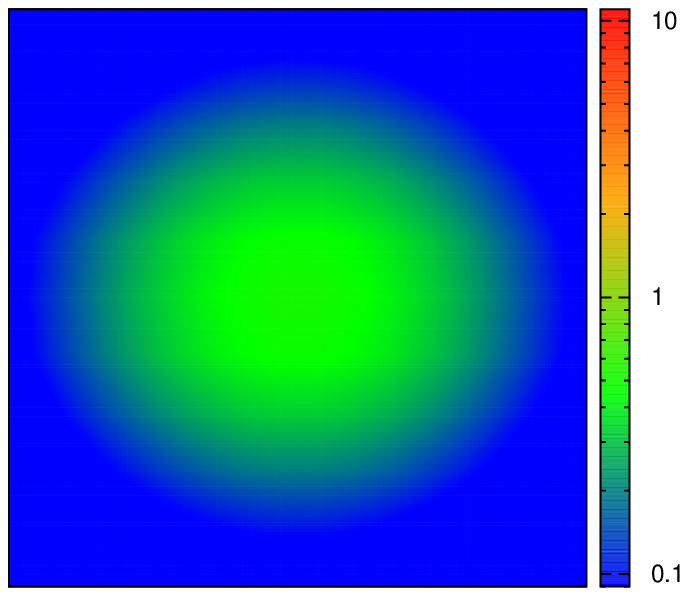}
\ec
\caption{\label{fig:fig1}
Time-evolution of the column density $n_{2d}(x,z)=\int n(x,y,z)\,dy$ in 
the $x$-$z$-plane. The coordinates and $n_{2d}$ are shown in arbitrary 
units, cf.~\cite{Bluhm:2015raa,Schafer:2010dv} for details. Snapshots are 
taken at times $t/t_0=0, 2, 8$ in units of the inverse geometric mean of 
the trap frequencies $t_0$.}
\end{figure}
Figure~\ref{fig:fig1} shows, as an example, the simulation result for the time-evolution of the two-dimensional column 
density $n_{2d}(x,z)$ in case of a purely temperature-dependent shear viscosity. Initially, the gas cloud is an 
ellipsoid elongated in the $z$-direction. Pressure gradients accellerate the system preferably in the transverse plane 
($x$-$y$-direction), but viscosity counteracts this accelleration and slows it down considerably for large $\eta$. The 
transition region from fluid dynamical behavior to free streaming is density-dependent and shifts during the 
evolution. In contrast to the Navier-Stokes theory, however, anisotropic fluid dynamics does not break down in the 
dilute regions for a density-independent $\eta$. 

The framework can be confronted with measured expansion data to deduce the dependence of $\eta$ on $T$ and $n$. This 
was done in~\cite{Bluhm:2015bzi}. In~\cite{Bluhm:2015bzi}, we analyzed the expansion data of Cao et 
al.~\cite{Cao:2010wa} for temperatures in the range $0.79\leq T/T_F\leq 1.54$ with Fermi temperature $T_F$. 
This is far above the superfluid transition, $T_c\simeq 0.23\,T_F$, and the initial density profile can be 
approximated by a Gaussian. Scale invariance in the Fermi gas at unitarity implies that $\eta$ is given by a universal 
function of $T$ and $n$, $\eta=(mT)^{3/2}\cdot f\left(n/(mT)^{3/2}\right)$. In~\cite{Bluhm:2015bzi}, we verified 
this form with the ansatz $\eta=\eta_0(mT)^{3/2}\left(mT/n^{2/3}\right)^c$ and found that the expansion data is 
best described with an exponent $c$ consistent with zero. This implies that in the high-$T$ limit $\eta$ depends only 
on temperature as predicted by kinetic theory, where the best fit was obtained for $\eta=0.282(mT)^{3/2}$ which agrees 
impressively with the kinetic theory expectation 
$\eta_{\rm CE}\simeq 0.269(mT)^{3/2}$~\cite{Bruun:2005,Bruun:2006,Schaefer:2014xma}. 

\section{Shear viscosity away from the unitarity limit}%

In the unitarity limit, the bulk viscosity $\zeta$ vanishes. For finite $s$-wave scattering length $a$, however, scale 
invariance is lost and $\zeta\ne 0$. In~\cite{Schaefer:2013oba}, this behavior was studied within kinetic theory for 
the dilute Fermi gas. It was found that $\zeta$ depends quadratically on the conformal symmetry breaking parameter, 
$z\lambda/a$, where $z$ is the fugacity and $\lambda$ the thermal wavelength of the fermions. In the high-$T$ limit, 
where fermionic quasiparticles are well defined, this can be understood as a medium effect due to a density-dependent 
fermion self-energy. To leading order in $a$, the ratio of bulk to shear viscosity reads 
$\zeta/\eta\sim\left(z\lambda/a\right)^2$. For the hot QGP, a comparable dependence of the viscosity ratio on the QCD 
measure for conformal symmetry breaking is found. In QCD, a non-zero bulk viscosity similarly arises from the scale 
breaking part in effective quark and gluon masses~\cite{Arnold:2006fz}. 

The scattering length dependence of the shear viscosity near unitarity was recently investigated in the expansion 
measurements by Elliott et al.~\cite{Elliott:2014nn}. For low $T$, an anomalous shift of the minimal shear viscosity 
away from unitarity toward positive $1/a$ is observed, which disappears with increasing $T$. Qualitatively, this 
behavior can again be understood as a medium effect due to modifications of the fermionic dispersion relation and the 
collision integral in the Boltzmann equation. As dominant effect, Pauli-blocking in the scattering amplitude can be 
identified. A systematic fugacity expansion in the dilute regime yields near unitarity~\cite{Bluhm:2014uza} 
\be
\label{equ:Expansion}
 \eta = \eta_\infty\left(1 + c_0 \left(\frac{\lambda}{a}\right)^2 + c_1 \left(\frac{z\lambda}{a}\right) 
 + \dots \right)
\ee
with $c_0=1/(4\pi)$, $c_1\simeq -0.03325$ and $\eta_\infty$ the shear viscosity at unitarity. At leading order, $\eta$ 
is independent of $z$ and even in $a$ and the minimum is reached at unitarity, while the next-to-leading order 
correction becomes more important with decreasing $T$, shifting the minimum toward $a>0$, i.e.~to the BEC-side of the 
Feshbach resonance. This implies that Pauli-blocking is more efficient on the BCS-superfluid side. Away from the 
unitarity limit, the ratio $\zeta/\eta$ shows an interesting dependence on the conformal symmetry breaking parameter 
which can be compared to the behavior found for the QGP near the confinement transition~\cite{Bluhm:2011xu}. 

\section{Conclusions}%

Cold Fermi gases can form strongly coupled quantum fluids under experimentally controllable conditions. This provides 
a unique opportunity to study certain properties of other strongly coupled quantum systems like the QGP. Almost perfect 
fluidity, for example, is also seen in expansion measurements of cold Fermi gas clouds at and near unitarity. 
Anisotropic fluid dynamics provides a realistic framework that allows us to reliably extract from these data the 
temperature and density dependence of the shear viscosity. Advances in kinetic theory showed, furthermore, that the 
observed anomalous shift of the minimal shear viscosity near unitarity can be understood qualitatively from microscopic 
calculations for the dilute gas limit. 

\vspace{2mm}
{\it Acknowledgments}\,: This work is supported in parts by the US Department of Energy under grant number 
DE-FG02-03ER41260. The work of M.~Bluhm is funded by the European Union's Horizon~2020 research and innovation 
programme under the Marie Sk\l{}odowska Curie grant agreement No 665778 via the National Science Center, Poland, 
under grant Polonez UMO-2016/21/P/ST2/04035. 


\end{document}